\documentclass{PoS}
\usepackage{epsfig}
\usepackage{amssymb}
\usepackage{amsfonts}
\usepackage{subfigure}

\title{Vortex topology and the continuum limit of lattice gauge theories}

\ShortTitle{Vortex topology and the continuum limit of LGT}

\author{\speaker{Giuseppe Burgio}\\
        Institut f. Theor. Physik, T\"ubingen\\
        E-mail: \email{burgio@tphys.physik.uni-tuebingen.de}}

\abstract{We study the stability of $\mathbb{Z}_2$ topological vortex 
excitations in d+1 dimensional $SU(2)$ Yang-Mills theory on the lattice at 
$T=0$. This is found to depend on d and on the coupling considered. We discuss 
the connection with lattice artifacts causing bulk transitions in the 
$\beta_A$-$\beta_F$ plane and draw some conclusions regarding the continuum 
limit of the theory.}

\FullConference{The XXV International Symposium on Lattice Field Theory\\
		 July 30 - August 4 2007\\
		 Regensburg, Germany}

\begin{document}

\section{Introduction}

The r\^ole of topological excitations in the effective mechanism for 
confinement for $SU(N)$ Yang-Mills theories has long been discussed in the 
literature, with Abelian monopoles \cite{'tHooft:1977hy} and $\mathbb{Z}_N$ 
magnetic vortices the \cite{'tHooft:1979uj} most popular candidates. 
For theories discretized on the lattice in the fundamental representation
a large number of studies is available on this subject (see e.g. these 
proceedings).

On the other hand, being $SU(N)/\mathbb{Z}_N$ the actual gauge group of
continuum {\it pure} Yang-Mills theories and resting the appearance of
$\mathbb{Z}_N$ vortices excitations indeed on such invariance 
\cite{'tHooft:1979uj}, it is surprising how scarce the lattice literature 
analyzing topological mechanism of confinement in adjoint discretizations is. 
A partial excuse for such failing can be sought in the difficulties connected 
to any numerical study of $SU(N)/\mathbb{Z}_N$ on the lattice. In 3+1 
dimensions the theories exhibit in the $\beta_A$-$\beta_F$ phase diagram
bulk transitions \cite{Greensite:1981hw,Bhanot:1981eb} (see Fig.~\ref{plane})
related to the condensation of $\mathbb{Z}_N$ magnetic monopoles $\sigma_{c}$ 
and electric vortices $\sigma_{l}$ \cite{Halliday:1981te,Halliday:1981tm}.

In a series of papers \cite{Barresi:2001dt,Barresi:2002un,Barresi:2003jq,%
Barresi:2003yb,Barresi:2004qa,Barresi:2004gk,Burgio:2005xe,Barresi:2006gq,%
Burgio:2006dc,Burgio:2006xj} such gap was filled for $SO(3) \simeq 
SU(2)/\mathbb{Z}_2$, leading to two important results. On one hand the 
connection of bulk transitions with the stability of $\mathbb{Z}_2$ magnetic
vortices was established, i.e. as expected \cite{Lubkin:1963,Coleman:1982cx}
well defined magnetic $\mathbb{Z}_2$ topological sectors have been found to 
exist only where $\mathbb{Z}_2$ magnetic monopoles cease to condense. On the 
other hand the $\mathbb{Z}_2$ magnetic vortex free energy was found to behave
quite differently as naively expected.

Although the ultimate goal is a throughout analysis of the latter result
for any $N$ and dimension, we concentrate here at first on the former to
see to what extent it can be extended to the whole $\beta_A$-$\beta_F$ phase
diagram in various dimensions. This is a necessary precondition for any 
future meaningful analysis of the $\mathbb{Z}_N$ magnetic vortex free energy.
We will show preliminary results indicating that the $T=0$ phase diagram as 
obtained from $\mathbb{Z}_2$ magnetic and electric vortices has a richer 
structure than previously believed by looking at $\mathbb{Z}_2$ magnetic 
monopoles. Although a similar picture emerges for $1+1$ and $3+1$ 
dimensions and for higher $N$ as well, we will concentrate here on 
$N=2$ in $2+1$ dimensions. A full analysis will appear in short time 
\cite{full}.

\section{Action and observables}

\begin{figure}[htb]
\begin{center}
\includegraphics[width=6.5cm]{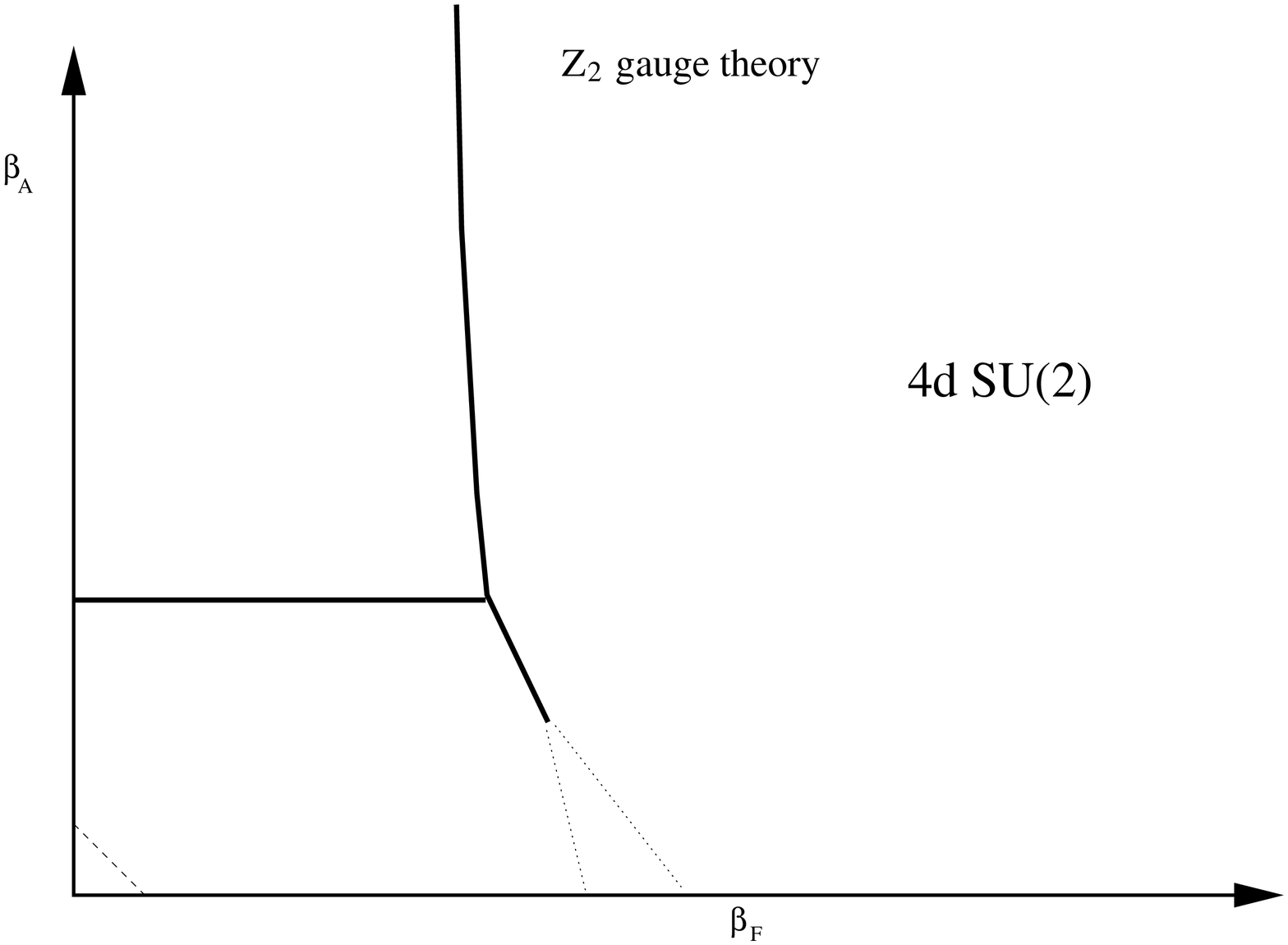}
\includegraphics[width=6.5cm]{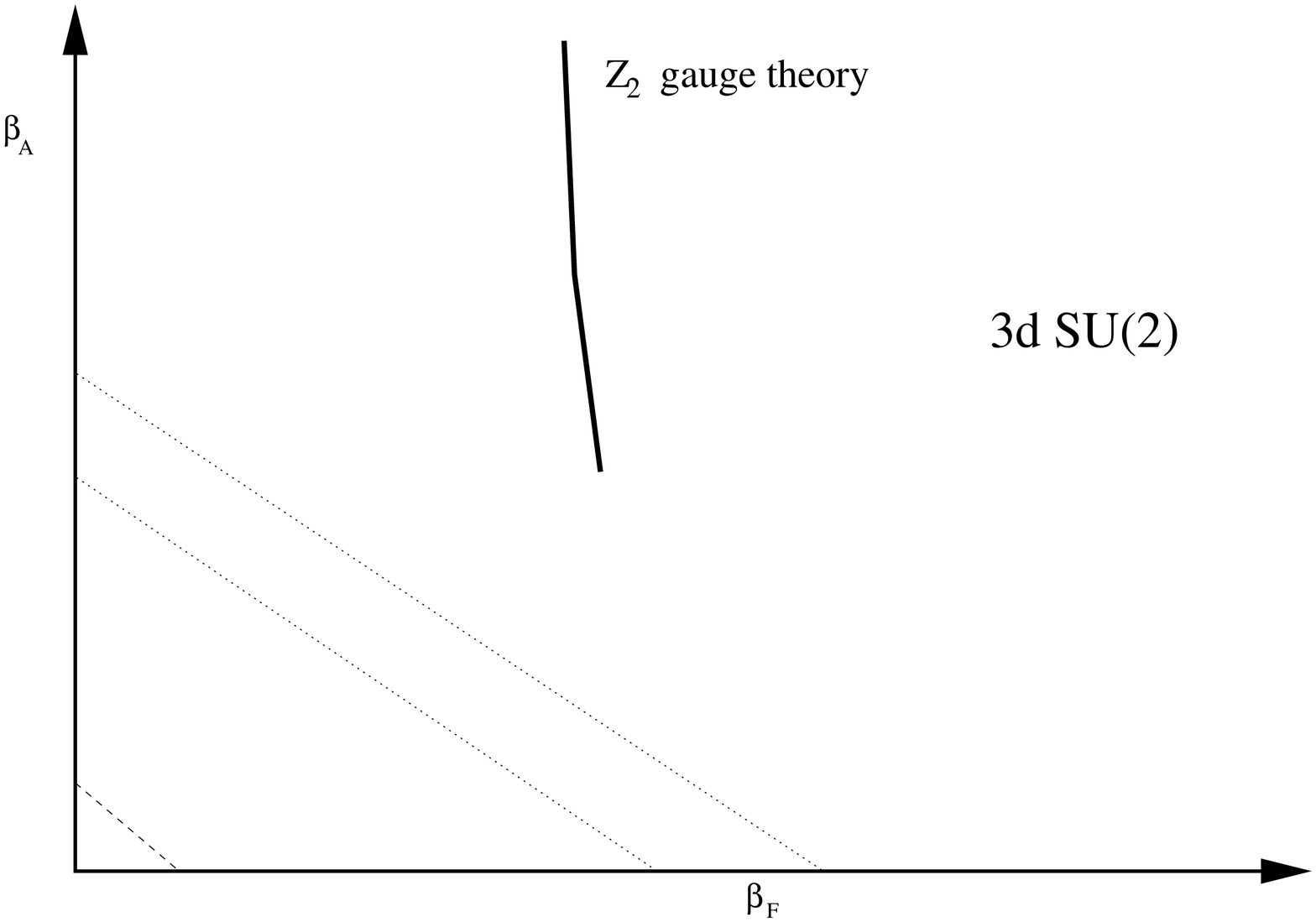}
\end{center}
\caption{Phase diagram from $\sigma_{c}$ and $\sigma_{l}$ of the 
$\beta_A$-$\beta_F$ plane in 3+1 (left) and 2+1 (right) dimension. Straight 
lines are bulk transitions, dotted lines show the crossover regions and dashed
lines the roughening transitions.}                             
\label{plane}
\end{figure}

We will consider the $N=2$ mixed fundamental-adjoint Wilson action in $d+1$ 
dimensions:
\begin{equation}
S = \beta_{A}\sum_{P}\Bigg(1-\frac{1}{3}Tr_{A}U_{P}\Bigg)+\beta_{F}\sum_{P}
\Bigg(1-\frac{1}{2}Tr_{F}U_{P}\Bigg)\,;\;\;
\frac{1}{g^2} =  \frac{1}{4}\beta_F+\frac{2}{3}\beta_A\,.
\label{mixed}
\end{equation}
Fig.~\ref{plane} shows the common picture as obtained in
\cite{Bhanot:1981eb,Baig:1987qa} for $d=2,3$. Order parameters for the 
transitions/crossovers (except for the roughening one \cite{Drouffe:1980dp}) 
are $\mathbb{Z}_2$ magnetic monopole $\sigma_{c}$ and electric vortices 
$\sigma_{l}$ densities $M$ and $E$ \cite{Halliday:1981te,Halliday:1981tm}:
{\setlength\arraycolsep{1pt}
\begin{eqnarray}
M&=&1-\langle\frac{1}{N_{c}}\sum_{c}\sigma_{c}\rangle\hspace{2cm}\sigma_{c}=
\prod_{P\epsilon\partial c}\sigma_{P}\;\;\;\in {SO(3)}\label{mmon}\\
E&=&1-\langle\frac{1}{N_{l}}\sum_{l}\sigma_{l}\rangle\hspace{2cm}\sigma_{l}=
\prod_{P\epsilon\partial l}\sigma_{P}\;\;\;\;\in {SU(2)}\label{evor}
\end{eqnarray}}
Apart for the roughening 
transition, the $d=1$ dimensional phase diagram is believed to be trivial, 
since $\mathbb{Z}_2$ magnetic monopoles cannot exist and the 
$\mathbb{Z}_2$ gauge theory, whose behaviour Eq.~\ref{evor} reflects, is
trivial in 1+1 dimensions.

In the continuum {\it pure} Yang-Mills theories are known to allow large gauge
transformations linked to $\pi_1(SU(N)/\mathbb{Z}_N) = \mathbb{Z}_N$, defining 
topological sectors corresponding to $\mathbb{Z}_N$ magnetic vortices 
\cite{'tHooft:1979uj}. These
are (evolving) points and lines for $d=2,3$ and instanton-like objects for 
$d=1$. On the lattice the fundamental discretization (quenched QCD) should only
allow one sector in the continuum limit, fixed by the periodic or twisted 
boundary conditions.
The adjoint theory with periodic boundary conditions is compatible with all 
sectors. In $SU(2)$ a suitable observable to measure the global $\mathbb{Z}_2$
magnetic flux through the $\mu\nu$ plane is given by
\begin{equation}
z_{\mu\nu} = \frac{1}{L^{d-1}}\sum_{\vec{\rho}\bot{\mu\nu}} \prod_{x \in 
{\mu,\nu} {\rm plane}} 
{\sf sign}({\sf Tr}_{f}U_{\mu\nu}(x))
\end{equation}
Of course for the fundamental theory with periodic boundary conditions only
allows $z_{\mu\nu} = 1$ regardless of $d$ for $\beta_F \to \infty$. 

$\mathbb{Z}_N$ magnetic monopoles, which are particle like for $d=3$ and 
instanton-like objects for $d=2$, will spoil such picture, since 
they are source of {\it open} $\mathbb{Z}_N$ magnetic vortices 
\cite{Lubkin:1963,Coleman:1982cx}. Since to any {\it abelian} monopole of 
charge $k$ corresponds a $\mathbb{Z}_N$ monopole of charge ${\rm mod}_N(k)$
in the continuum limit, where the latter ought not to exist, the former 
are only allowed charges $k \propto N$, i.e. only these are compatible with 
closed $\mathbb{Z}_N$ magnetic vortices \cite{Lubkin:1963,Coleman:1982cx,%
Frohlich:2000zp}, as the case in pure $SO(3)$ approaching the continuum limit 
indicates \cite{Burgio:2006xj}. This also implies that in a 
$\mathbb{Z}_N$ monopole background stable vortices are submerged by the 
related open vortex background and 
$z_{\mu\nu}$ will have no well defined single value through all parallel 
$\mu\nu$ planes, averaging to $z_{\mu\nu}=0$ for any MC configuration, 
as indeed shown in \cite{Burgio:2006dc,Burgio:2006xj}. Along 
the known bulk transition lines the order parameter
\begin{equation}
z=\frac{2}{d(d+1)} \sum_{\mu\nu} \langle |z_{\mu\nu}|\rangle 
\end{equation}
follows $M$, their behaviour being indistinguishable, so that in the strong 
coupling region $z=0$ while as the
theory approaches the continuum limit $z_{\mu\nu} \to \pm 1$ and $z \to 1$. 
What happens however across the crossover regions? Does $z$ approach its 
asymptotic value following $M$ and $E$? And in 1+1 dimensions, where 
no monopoles can appear, but the boundary conditions still dictate the global
$\mathbb{Z}_N$ magnetic flux, so that the appearance of the wrong one can
still characterize the strong coupling regime? We concentrate on $\beta_A=0$ 
for $d=2$ as an example.

\section{Results}

\begin{figure}[htb]
\begin{center}
\includegraphics[width=9cm]{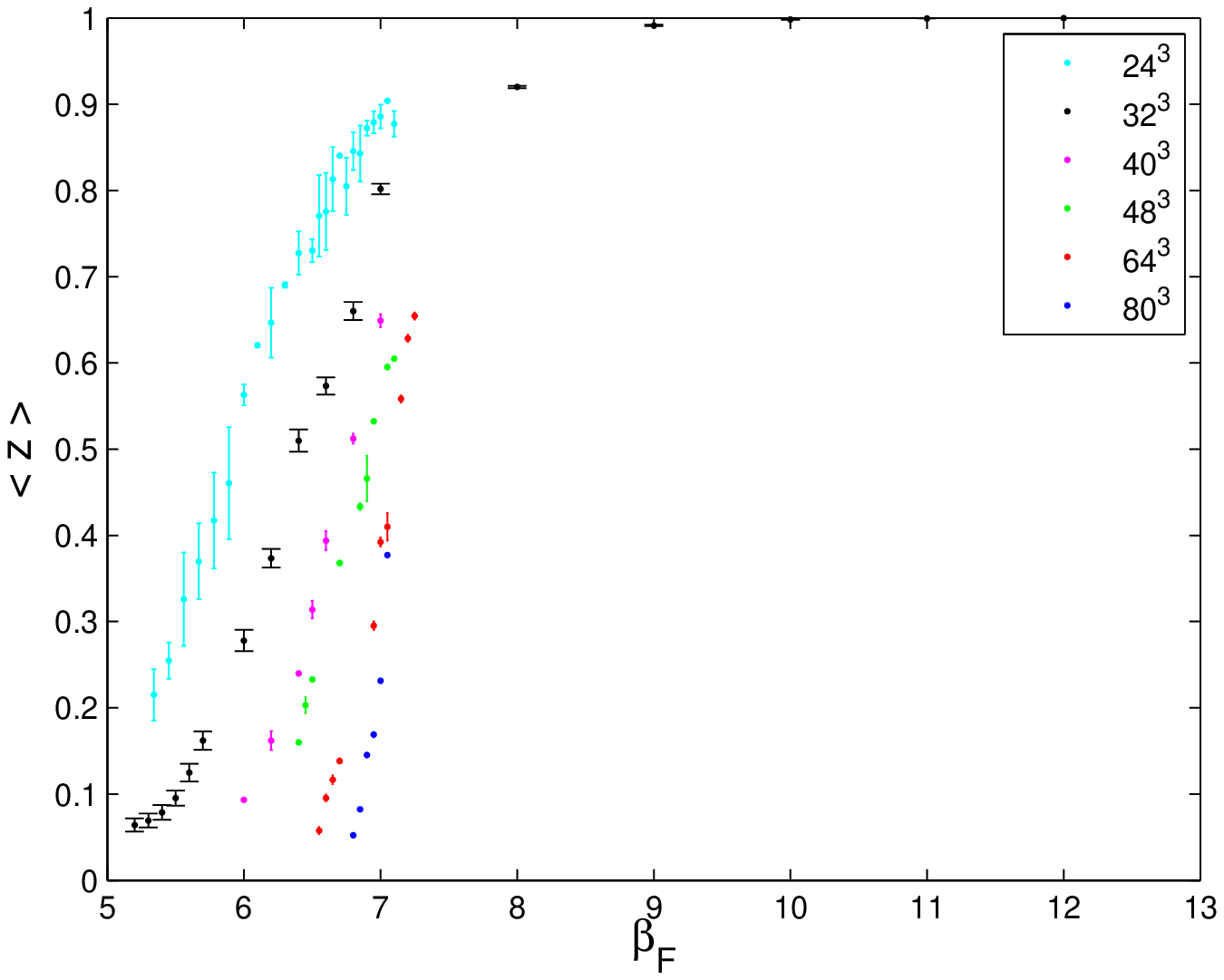}
\includegraphics[width=9cm]{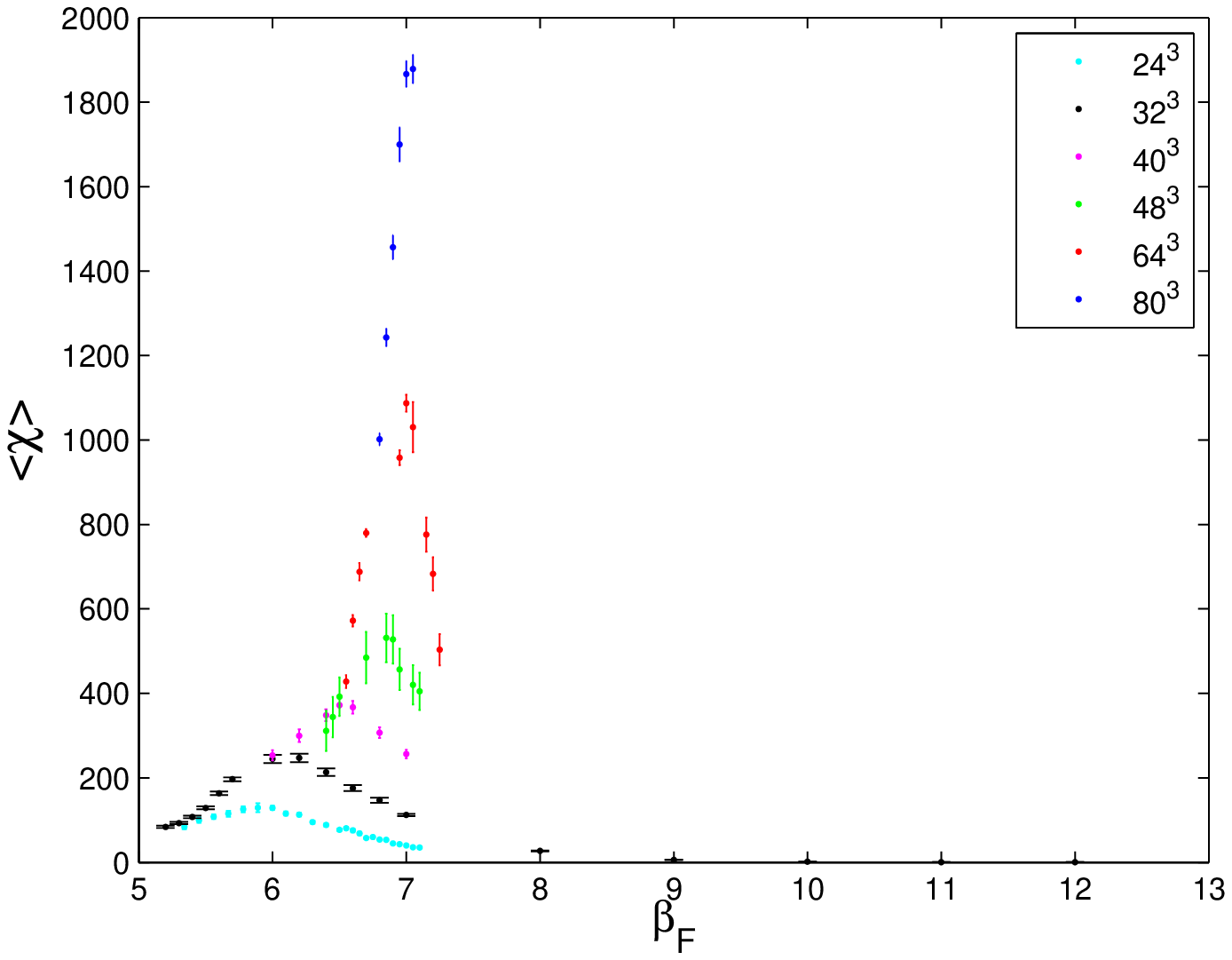}
\end{center}
\caption{Order parameter $z$ (left) and its susceptibility $\chi$ for different
volumes.}                             
\label{par}
\end{figure}

Fig.~\ref{par} shows $z$ and its susceptibility $\chi$ for increasing volume 
as a function of $\beta_F$ at $T=0$ in 2+1 dimensions.
We remind that along the crossover $M$ and $E$ peak around $\beta_F =4 - 5$. 
The critical behaviour of $z$ at higher $\beta_F$ is evident. A similar 
picture emerges also for $\beta_A \neq 0$, for $N=3$ and for $d=1$ and 3 
\cite{full}. Establishing the 
properties of the transition is however a hard task. Integrated 
autocorrelations for $z$ show a strong critical slowing down approaching 
criticality. Moreover a direct investigation of 
the plaquette and the specific heath shows no sign of critical behaviour, 
excluding a 1$^{\rm st}$ or standard (i.e. divergent) 2$^{\rm nd}$ order 
transition. 
\begin{figure}[htb]
\begin{center}
\includegraphics[width=9cm]{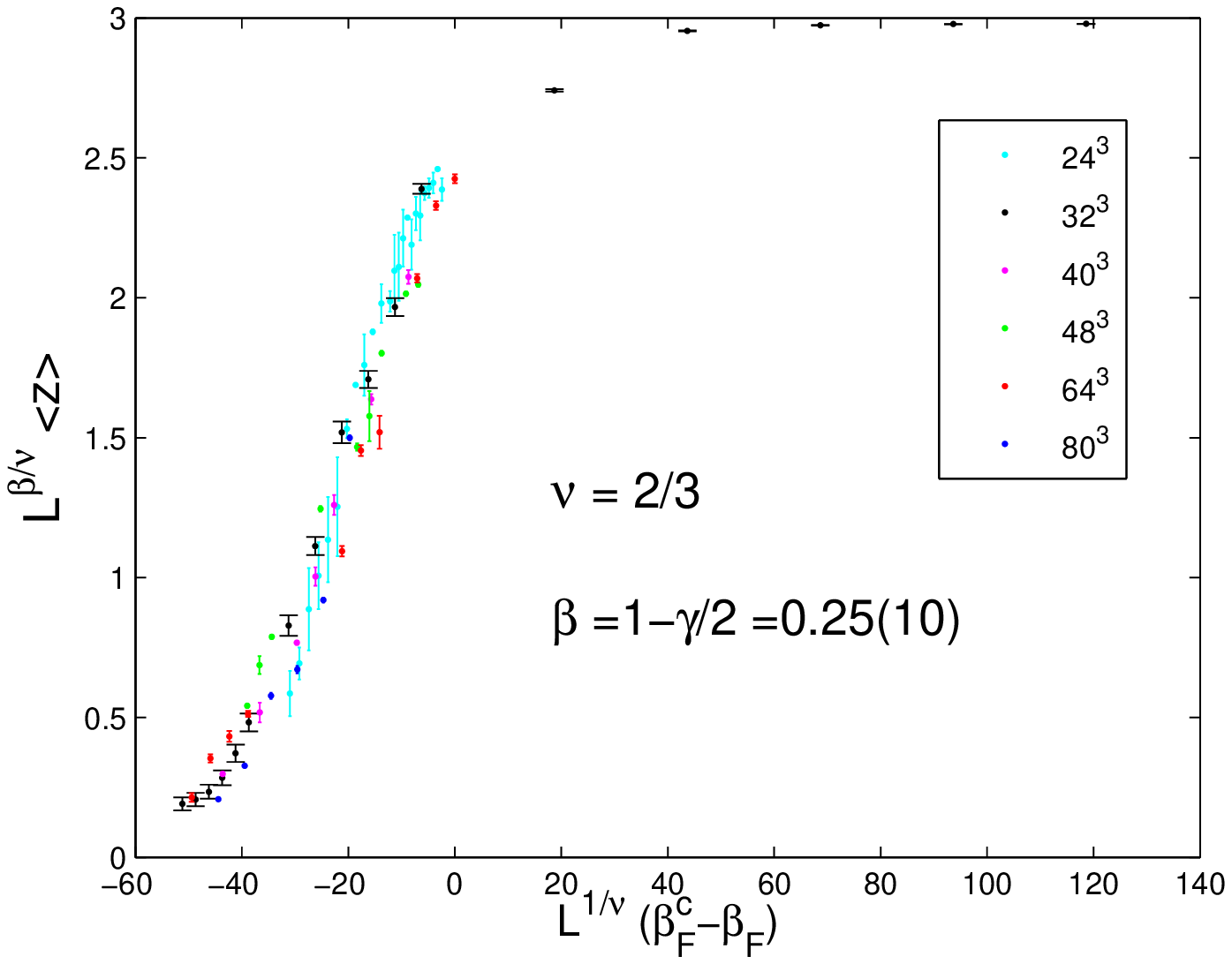}
\includegraphics[width=9cm]{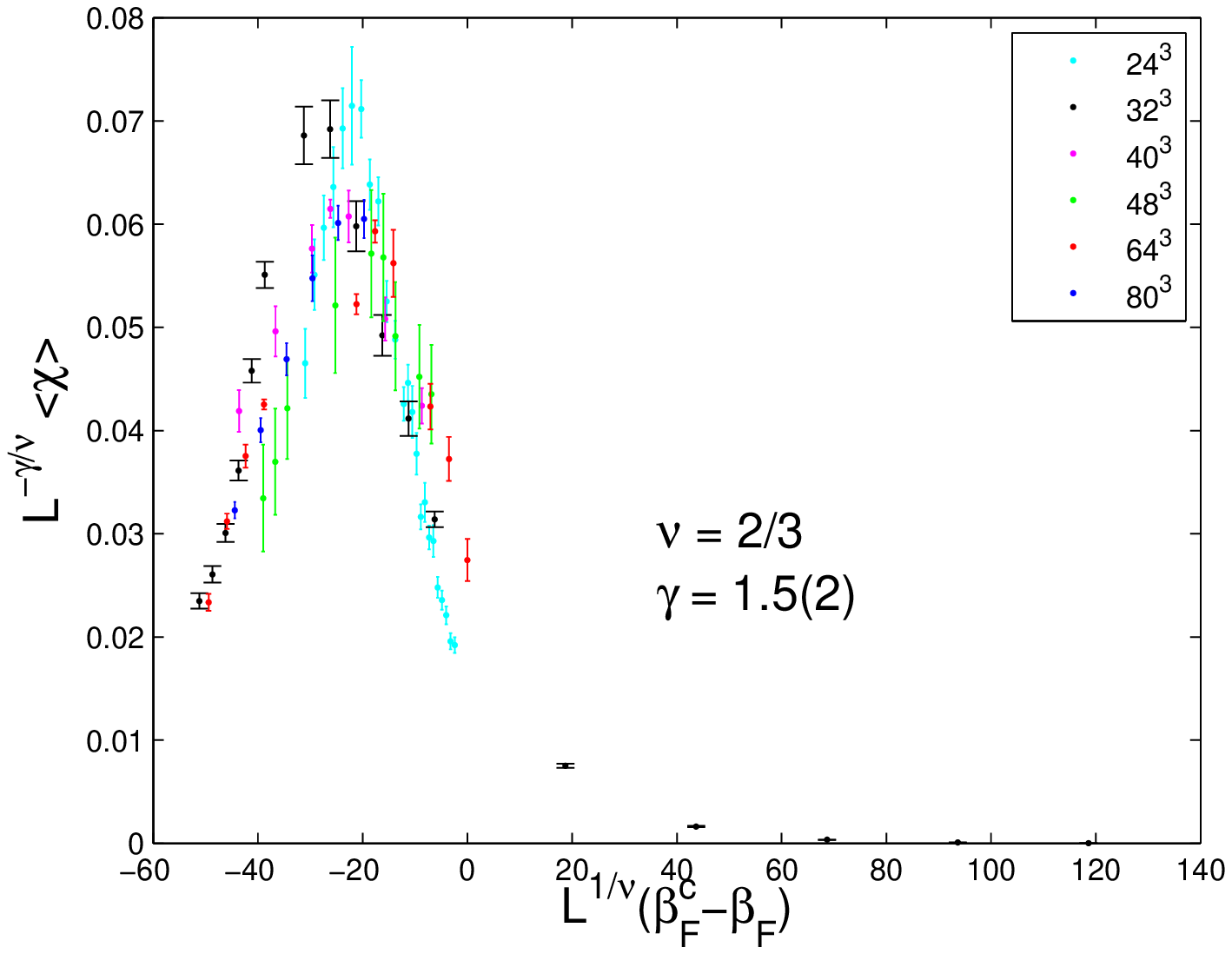}
\end{center}
\caption{FSS for $z$ (left) and its susceptibility $\chi$ for different
volumes.}                             
\label{ffs}
\end{figure}

This leaves however room for a discontinuous 2$^{\rm nd}$ order
with a small gap in the specific heath or higher ($\geq 3^{\rm rd}$) 
transition. Indeed this would be not unheard of, since e.g. a third order 
transition is know to exist in $d=1$ in the large $N$ limit and conjectured 
to extend also to higher dimensions \cite{Gross:1980he}. Following standard
techniques we check the consistency of FSS assuming the ``critical 
exponents'' extracted from hyperscaling relations for a 2$^{\rm nd}$ order 
discontinuous transition \cite{Janke:2006wr}, i.e. $\nu = 2/3$, $\gamma = 4/3$
and $\beta = 1/3$. Actually we fit $\gamma = 1.5(2)$ from $\chi_{\rm max}(L) 
\simeq L^{\gamma/\nu}$. The curves obtained by tuning $\beta_F^{\rm c} = 
7.3(1)$ and using $\beta = 1-\gamma/2 = 0.25(10)$ are shown in Fig~\ref{ffs}. 
Given the 
high systematic errors coming from autocorrelations $\gamma$ and $\beta$ are
in good agreement with the theoretical prediction. This FSS analysis should
however be considered still tentative pending better precision in the
data and alternative independent methods to establish the order of the 
transition. A direct analysis of Fisher zeroes \cite{Janke:2006wr} seems the 
most promising.

\section{Conclusions}
\begin{figure}[htb]
\begin{center}
\includegraphics[width=6.5cm]{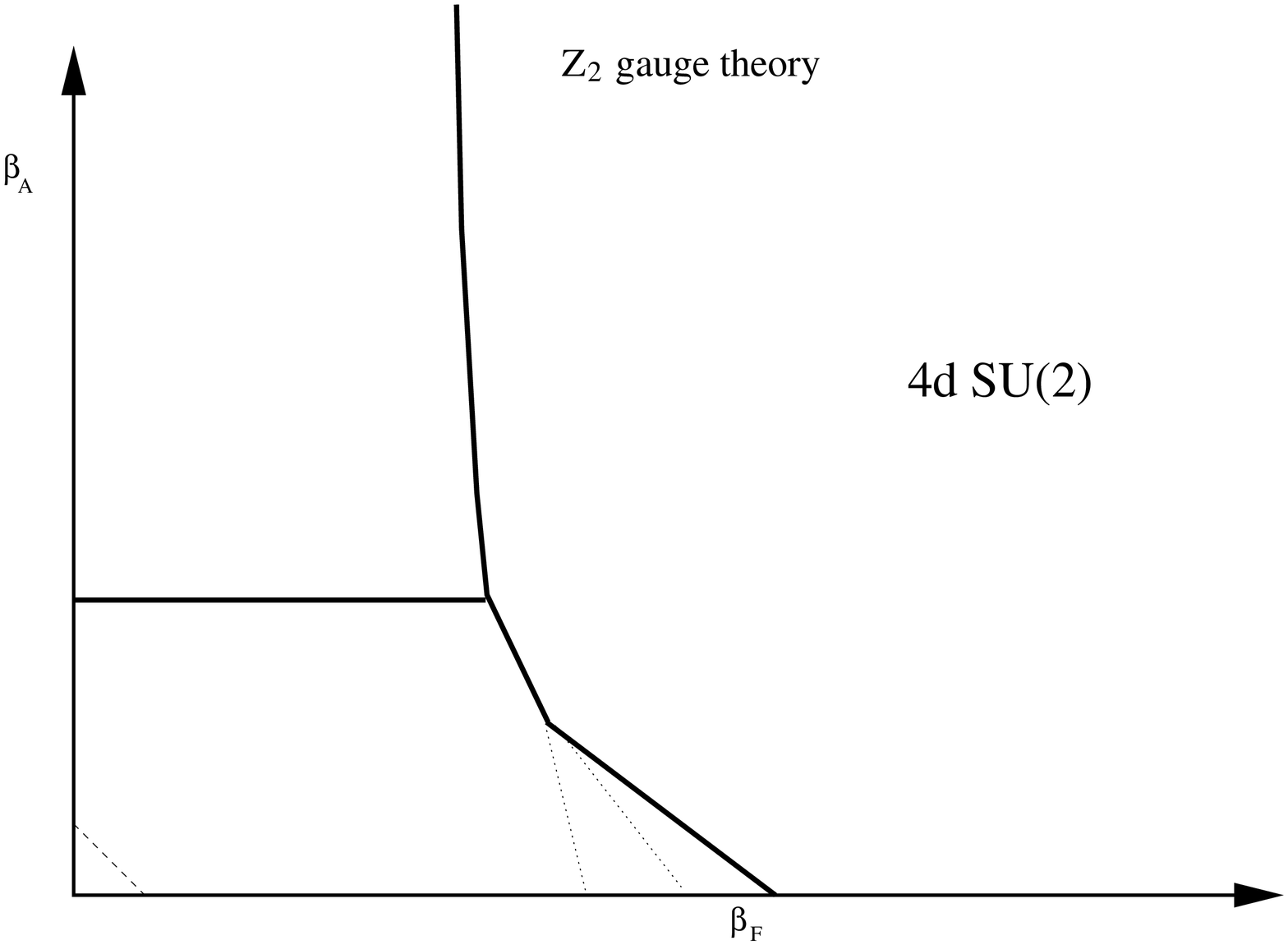}
\includegraphics[width=6.5cm]{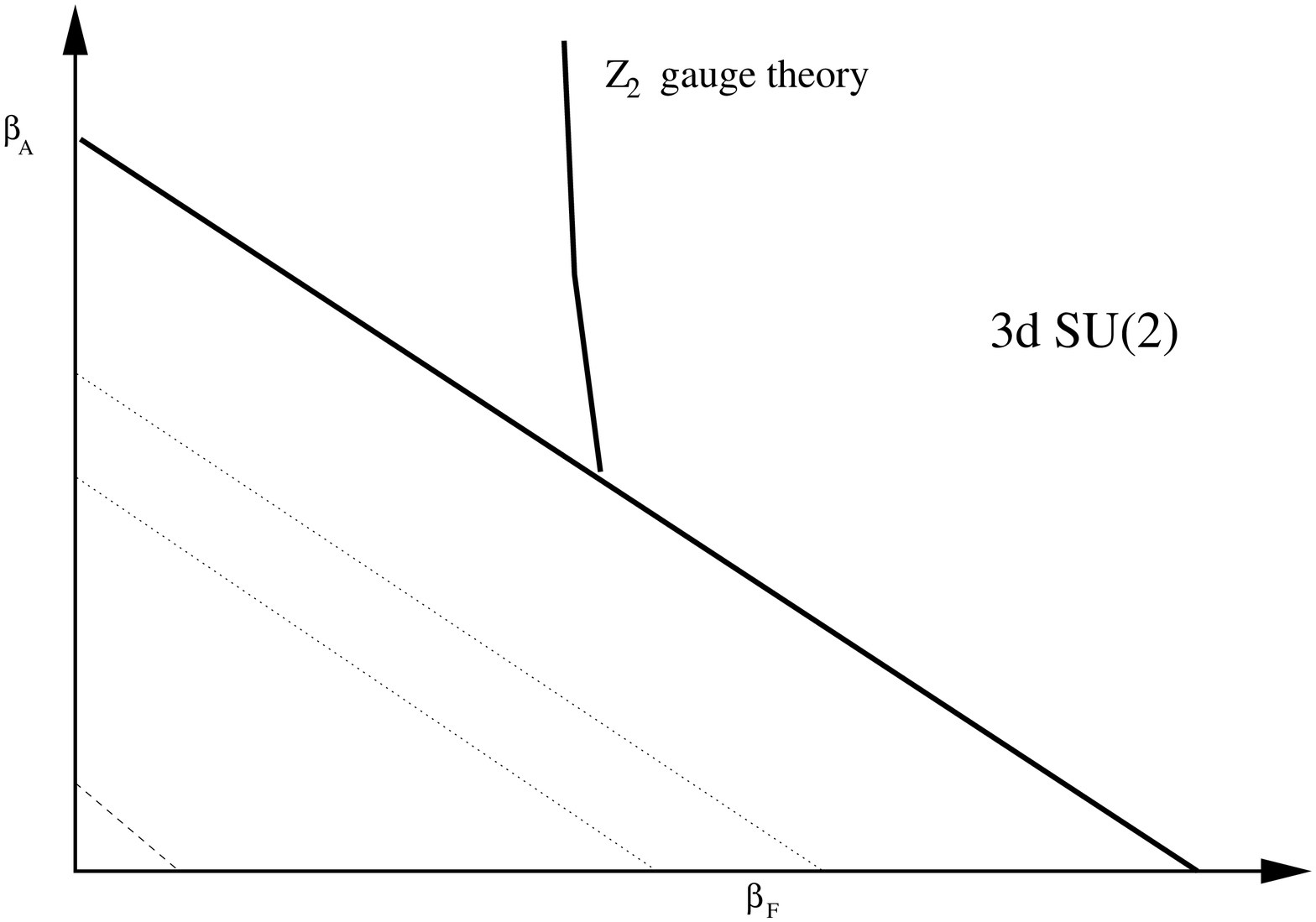}
\end{center}
\caption{Phase diagram from $z$ and $\sigma_{l}$ of the 
$\beta_A$-$\beta_F$ plane in 3+1 (left) and 2+1 (right) dimension. Straight 
lines are bulk transitions, dotted lines show the crossover regions and dashed
lines the roughening transitions.}                             
\label{plane_new}
\end{figure}

We have given evidence that in $2+1$ dimension the pure $SU(2)$ theory 
undergoes a bulk-like transition at $\beta_F = 7.3(1)$. The order parameter
of such transition $z$ measures the stability of $\mathbb{Z}_2$ magnetic
vortices. A preliminary FSS analysis shows the critical exponents to be 
consistent with a 2$^{\rm nd}$ order discontinuous transition. A similar
pictures emerges for $\beta_A > 0$, other spacial dimensions and $N>3$. 
Fig~\ref{plane_new} shows how e.g. the 3+1 and 2+1 $\beta_A$-$\beta_F$ phase 
diagram would indeed look like if such results should be confirmed.
In 1+1 dimensions a phase transition line would also separate the strong from
the weak coupling phase in the whole $\beta_A$-$\beta_F$ plane.

The preliminary character of the results only concerns their quantitative
analysis, i.e. extracting the exact position, order and critical exponents
of the transitions in the $\beta_A$-$\beta_F$ plane for different dimensions.
To this goal a better control of autocorrelations and the use of methods 
alternative to those here exposed would be most welcome. In particular, 
studying the density of Fisher zeroes from the complexified partition function 
\cite{Janke:2006wr} should provide a direct independent method to establish
$\beta^{\rm c}$ and the order and type of the transition. 
Nevertheless, the presence of a critical behaviour for $z$ is evident from the
data and needs to be addressed independently of the above caveats.

Although the presence of such transitions, being most likely of high order, 
might not be evident in most observable, it might affect
in principle any attempt to use RG-flow methods to connect weak to strong
coupling regions, especially if basing on vortex related observables 
\cite{Tomboulis:2007iw}. To this goal it would be important to establish 
whether the transition line has an end point at $\beta_A < 0$ and how it does 
connect to the bulk lines detected through $\sigma_l$. 

Moreover any
result on the r\^ole of topological excitations for confinement,
abelian monopoles and $\mathbb{Z}_N$ magnetic vortices in particular, 
should be indeed rechecked above the transition lines in the region 
connected to the continuum theory, where both can take their correct value. 
Being the results found in the only case where this was done 
\cite{Burgio:2006dc,Burgio:2006xj} somehow surprising, although partly expected
and partly justifiable through analysis of the Hilbert space states 
\cite{Burgio:1999tg}, crosschecks and better understanding would be most 
welcome.

\section*{Acknowledgements}
We wish to thank R.~Kenna for stimulating discussions and directing us to the 
relevant literature on Fisher zeroes and higher order phase transitions. A 
special thank goes to F.~Bursa for stimulating correspondence on the 1+1 
dimensional theory. We 
also would like to thank H.~Reinhardt and M.~Quandt for interesting 
discussions. This work was supported by DFG grants Re 856/4-1,2,3 and 
Re856/5-1.

\end{document}